# An Observational Study of Engineering Online Education During the COVID-19 Pandemic

**Shadnaz Asgari**[1,2], **Jelena Trajkovic**[2], **Mehran Rahmani**[3], **Wenlu Zhang**[2], **Roger C. Lo**[4], **and Antonella Sciortino**[3,5]

[1]Department of Biomedical Engineering, California State University, Long Beach, California, USA
[2]Department of Computer Engineering and Computer Science, California State University, Long Beach, California, USA
[3]Department of Civil Engineering and Construction Engineering Management, California State University, Long Beach, California, USA
[4]Department of Chemical Engineering, California State University, Long Beach, California, USA
[5]College of Engineering, California State University, Long Beach, California, USA

Corresponding author: Shadnaz Asgari (e-mail: Shadnaz.Asgari@csulb.edu).

This research is partially supported by CSULB Champions program through Coronavirus Aid, Relief, and Economic Security (CARES) act funding.

**ABSTRACT** Although online education has become a viable and major component of higher education in many fields, its employment in engineering disciplines has been limited. COVID-19 pandemic compelled the global and abrupt conversion of conventional face-to-face instruction to the online format. The negative impact of such sudden change is undeniable. Urgent and careful planning is needed to mitigate pandemic negative effects on engineering education, especially for vulnerable, disadvantaged, and underrepresented students who have to deal with additional challenges (e.g. digital equity gap). To enhance engineering online instruction during the pandemic era, we conducted an observational study at California State University, Long Beach (a minority-serving institution). 110 faculty and 627 students from six engineering departments participated in our surveys and answered quantitative and qualitative questions to highlight the challenges they experienced during the online instruction in Spring 2020. In this work, we present the results of these surveys in detail and propose solutions to address the identified issues including logistical, technical, learning/teaching challenges, assessment methods, and hands-on training. As the pandemic continues, sharing these results with other educators can help with more effective planning and choice of best practices to improve the online engineering education during COVID-19 and beyond.

## I. INTRODUCTION

Engineering education has been traditionally content-centered, hands-on, design-oriented, and focused on the development of critical thinking or problem-solving skills [1]. Various teaching methodologies have shown efficacy in enhancement of engineering education including active learning [2], flipped classroom [3] and project-based learning [4-6].

Over the last decade, online education has become a viable and major component of higher education in many fields including business administration and management, psychology, criminal justice, etc. On the other hand, online education is not yet widespread throughout all engineering disciplines [1] but it is limited to few carefully selected courses in each engineering discipline or mainly at the master's and postgraduate levels [7]. Hands-on training to work with equipment, instruments and materials in controlled laboratory setting is an inherent and necessary aspect of a successful engineering education. However, addressing this essential aspect within a fully online teaching platform is challenging. Moreover, many students prefer to learn difficult concepts face-to-face [8] and believe that the online education might not be the best choice for a deep level of learning [9]. Converting a course from conventional face-to-face to online format is also time consuming and requires the instructor's familiarity with (or willingness to learn about) online learning or instructional tools including the learning management system (LMS). Another issue is the difficulty of designing fair and yet rigorous online assessments to minimize cheating and plagiarism [10]. A successful engineering education requires creating and maintaining a reliable and robust infrastructure that supports both faculty and students [7, 11].

In an unprecedented event, the COVID-19 pandemic caused the sudden temporarily closure of most educational institutions around the world and consequently, the inevitable conversion of their conventional face-to-face instruction into fully online (or blended/hybrid) format in a short transitional time. According to UNESCO statistical data, since the onset of pandemic, more than 1.5 billion students worldwide (90.1% of total enrolled learners) have been affected by the COVID-19 closures and subsequent educational changes [12].

Urgent and careful planning is needed to mitigate the negative impact of the pandemic changes on engineering education especially for vulnerable, disadvantaged and underrepresented students facing substantial challenges beyond their academic responsibilities including family obligations, financial burden and additional employments [13-15]. Additional efforts shall be also taken to guarantee that the online instruction of engineering courses still meet the rigorous requirements of the program accreditations, e.g. ABET.

California State University, Long Beach (CSULB) is one of the largest four-year and most diverse universities in the U.S. Approximately 52% of CSULB student body are NSF-defined underrepresented minority including 59.2% female, 46.9% Hispanic, 4.5% African American and 1% Native American [16]. Also, more than half of our students are low-income or first-generation college students. As a result, CSULB is recognized as a minority serving institution: namely Hispanic, Asian American, Native American, and Pacific Islander-Serving Institution. Due to COVID-19, majority of educational programs in CSULB were converted to fully online format within a transitional period of 10 days in March 2020. This included all the programs offered by the College of Engineering (COE) that consists of six departments, with more than 250 faculty and 5000 students. The unprecedented circumstances of global pandemic enforced the swift conversion of the mode of instruction. Hence, the teaching materials and assessment methods had to be developed "on the fly". Both our students and faculty encountered various challenges during the online instruction in Spring 2020. By the end of the semester in May 2020, CSULB announced that Fall 2020 semester was going to be in the alternative mode of instructions, as well. Thus, the majority of the courses were scheduled to be offered in synchronous online format (where faculty and students are expected to attend the live online sessions). Few classes, where the face-to-face component is considered essential to meet the learning outcomes of the course and therefore could not be conducted fully online, (e.g. laboratories and senior design capstone projects) were exempted and offered in hybrid/blended format.

Following this announcement, a team of selected faculty from six engineering departments started working on identifying teaching/learning obstacles, challenges, and opportunities for improvement to better plan for the online instruction in Fall 2020. Sloan's online learning consortium has defined the five pillars of high-quality online education as: learning effectiveness, student satisfaction, faculty satisfaction, access, scale and cost [1]. Given these factors, our team decided to take a systematic approach for improvement of the instruction by first conducting faculty and student surveys and then proposing interventions to address the identified issues. This paper presents the results of our surveys and provides a summary of proposed solutions. Sharing these results with other educators in engineering field can facilitate a more robust continuity of education during ongoing pandemic. It can also aid with overall improvement of online engineering education in the post-pandemic era especially for those universities with a large percentage of minority and first-generation/low-income students.

## II. MATERIALS AND METHODS

### A. ENGINEERING EDUCATION AT CSULB

CSULB COE currently serves more than 5000 undergraduate and graduate students through a total of 11 programs hosted by six departments: Biomedical Engineering (BME), Chemical Engineering (CHE), Civil Engineering & Construction Engineering Management (CECEM), Computer Engineering & Computer Science (CECS), Electrical Engineering (EE), and Mechanical & Aerospace Engineering (MAE). COE offered 349 courses, for a total of 1004 sections in Spring 2020, and is currently offering 331 courses (5.5% hybrid and the rest fully online) in Fall 2020.

In 2010, CSULB started using an LMS called BeachBoard (BB)−developed and supported by the D2L (Desire 2 Learn) company as a customized version of their "Brightspace" platform. BB provides various features to facilitate the course instruction including a robust platform for communication between the instructor and students, sharing course materials with students, recording of lectures, discussion forums, design and management of assessments, assignments and grades. While some CSULB faculty members have employed (at least some of) BB features (e.g. gradebook) for their instruction on a regular basis, others might have opted out as its usage has not been mandatory.

Following the COVID-19 pandemic and during the short transitional period to online instruction, CSULB advised instructors to mainly focus on learning/using BB (and Zoom videoconferencing) to convert their courses to the online format. This recommendation seemed reasonable given the availability and practicality of BB features.

### B. SURVEYS

Our goal was to identify and study the magnitude of various issues that our faculty and students encountered during our six weeks of online instruction in Spring 2020 (March 23-May 8) and plan for an enhanced online instruction in Fall 2020. The faculty and student surveys were designed holistically considering the overall verbal feedback received from stakeholders during the Spring 2020 online instruction. The faculty survey consisted of 10 multiple-choice and 2 free-response questions, while student survey included 6 multiple-choice questions with fill-in or additional comment options for each question.

The faculty survey questions covered a variety of online teaching issues including, but not limited to, the lack of access to necessary hardware (e.g. computer/tablet, scanner/printer, microphone/headset, camera), software and reliable internet connection. Some questions focused on various learning assessment methods that instructors used in Spring 2020 (or the ones they were planning to use in Fall 2020) including open-book or closed-book exams, synchronous or asynchronous exams, fully-online exam (using randomized questions on BB) or semi-online exams (where students solve the assigned problems on a paper, then scan and upload their solutions on BB). Some questions focused on proctoring exams and the instructors' perceived prevalence of cheating/plagiarism. We also asked faculty to indicate the topics that they were interested to enhance their skills on, e.g., basic or advanced BB features, Zoom features, automatic grading, etc. The two open-ended questions provided instructors additional opportunities to comment about their online teaching experience and make any suggestion or request to COE that could help with improvement of online instruction in Fall 2020.

The students survey was designed to identify the challenges students confronted during online instruction in Spring 2020, including lack of access to hardware, software, reliable internet connection, quiet/private space to study, potential issues of balancing study with work and family duties, and stress management. We also asked about difficulties students had during the synchronous classes on Zoom (e.g., lack of focus or engagement, instructor's lack of familiarity with technology) or during the online exams (e.g., time management, issues with methods of proctoring using camera). Copies of faculty and student surveys are enclosed in the appendix for the readers' further reference.

### III. RESULTS

The faculty survey was conducted using Qualtrics over a three-week period (June 20-July 10). Similarly, the student survey was designed and conducted in Qualtrics afterwards (July 27-August 12). This later timeframe was decided based on the assumption that more students (including the incoming students) might be available to participate in the survey closer to the beginning of the Fall 2020 semester (August 21). Participation in both surveys were anonymous.

A total of 110 instructors took the survey where 43% of them were full-time including tenured/tenure track faculties and the rest were part-time lecturers. Also, 627 students participated in the survey: First-year students (4%), Sophomore (14%), Junior (30%), Seniors (35%) and graduate students (17%). Fig. 1 shows the distribution of survey participants among various departments within the COE (question #1 on both surveys). We observe that all departments have relatively similar representations in terms of percentage of faculty and student participants (9% BME, 5-10% CHE, 15-23% CECEM, 19-22% CECS, 18-22% EE, and 21-26% MAE).

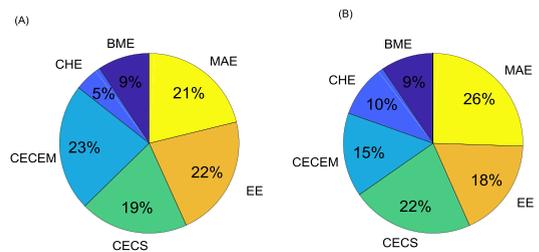

**FIGURE 1.** Distribution of the survey participants among various departments within the college of Engineering: (A) Faculty participants; (B) Student participants.

These percentages are consistent with the size of our departments in terms of number of faculty and students. This observation shows that our survey sample population could be a good representative of the general COE populations in terms of existing majors.

### A. LOGISTICAL CHALLENGES FOR BOTH STUDENTS AND FACULTY

Fig. 2 shows the percentages of survey respondents who indicated various logistical challenges they had during the online instruction period of Spring 2020 (question #2 on the faculty survey and question #3 on the student survey). Close to 15% of the faculty had issues with software license or no access to personal computer/tablet. About 20% of the faculty did not have access to microphone/headset or printer/scanner. 23% of faculty had no reliable internet connection, while 32% had no access to webcam or camera for the online instruction. Finally, 47% of the faculty indicated that they had no access or technical difficulties with online writing tools.

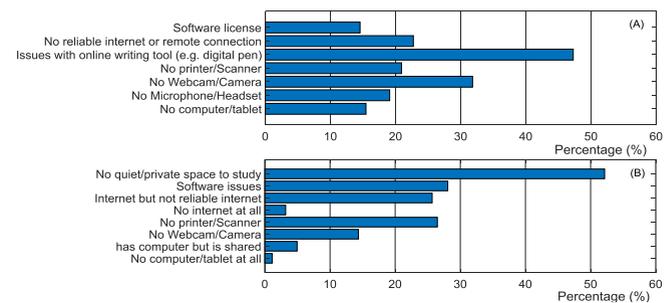

**FIGURE 2.** Logistical challenges of online instruction from perspectives of faculty and students. The horizontal access represents the percentage of survey participants who indicated the corresponding challenge. (A) Faculty respondents; (B) Student respondents.

Among the student respondents, 1% had no access to any computer/tablet, while close to 5% had only access to a shared computer at home. 3% had no internet connection, while 26% had issues with reliability of their internet. 28% indicated having issues with software access, while 26% had no printer/scanner at home.

## B. STUDENTS CHALLENGES WITH ONLINE INSTRUCTION

Fig. 3 summarizes the prevalence of challenges students had with online instruction during Spring 2020 (questions # 3-6 on the student survey). About 70% of students indicated difficulty in maintaining their focus or experiencing Zoom fatigue after attending multiple online sessions. 55% of students felt social disconnection from their classmates/peers, while 64% did not feel engaged during the online classes. 60% of the students felt there was a lack of clear guidance or communication from the instructors. Also, a quarter of students had issues with online submission of assignments and exams, mainly due to the lack of access to printer/scanner as we learned from students' optional comments.

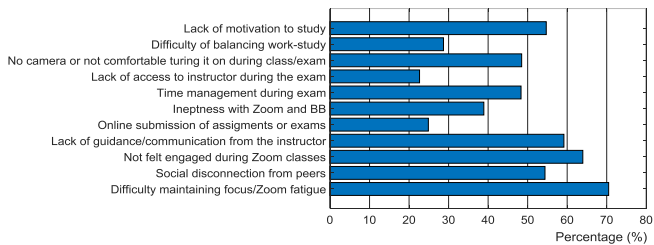

**FIGURE 3.** Prevalence of challenges students encountered during online instruction in Spring 2020.

About 40% of students had technical difficulty and ineptness issues with using or navigating through Zoom or BB. 48% of the students experienced time management issues during the online exams. In optional comments, some students expressed their frustration with not being able to go back to previous questions (a BB feature for the instructors to limit cheating). 23% of the students indicated that the unavailability of the instructor during the online exam (in contrast to in-person exam) caused challenges.

48% of the students specified that they either do not have camera or feel uncomfortable turning the camera/microphone on during the class or online exams (question #7 on the student survey). Optional comments revealed that many participants have privacy concerns with usage of camera/microphone or being recorded, especially if they were living in a crowded home or shared space. Furthermore, some students experienced an increased level of anxiety being watched on camera that hindered their focus and lowered their performance during the online exams. 28% of the students indicated that they had difficulty with balancing work and study. From the optional comments, we understood that the latter issue has been escalated for many during pandemic. Some parents had lost their jobs and consequently the whole family was relying on the part-time jobs of the younger adults (students) to survive financially.

Our survey also indicated that more than 50% of our students did not have access to a private or even quiet space to attend the online classes or to study. 55% of students also lacked motivation to study (question #3 on the student survey). The optional comments shed further light onto the lack of motivation. The uncertainty of the COVID-19 pandemic and loss of peer interaction/support were identified as the major contributing factors. Finally, 24% of the students rated their overall experience of online instruction (question #8 on the student survey) as satisfying, 37% found it dissatisfying while the rest (39%) were neutral.

## C. ASSESSMENT METHODS USED DURING ONLINE INSTRUCTION

Table 1 shows the prevalence of various methods that the faculty used to assess students' learning during the online instruction of Spring 2020. Semi-online refers to an exam where students solve the assigned problems on a paper, then scan and upload their solutions. Asynchronous exam refers to a take-home exam while a synchronous exam is the one conducted during the scheduled class or exam time. The survey allowed respondents to choose more than one assessment method per question (because faculty might have taught multiple classes and held more than one exam during the semester), thus the sum of the percentages would not equal to 100.

**TABLE I**
Learning assessment methods faculty used during the online instruction in Spring 2020. The respondents could choose more than one option for each question depending on the number of exams administered during the semester.

| Survey Question # | Assessment method | Percentage of faculty who employed the method |
|---|---|---|
| Question #3 | Fully online exam (e.g., BB quiz) | 63% |
|  | Semi-online Asynchronous exam | 28% |
|  | Semi-online Synchronous exam | 40% |
|  | project/term paper | 50% |
|  | Oral presentation/demo | 33% |
| Question #6 | Open-book/Open-note exam | 70% |
|  | Closed-book/Closed-note exam | 33% |

We observe that the fully online exams such as the BB quizzes were used by 63% of the faculty. BB quizzes provides the faculty with the convenient option of randomizing the order and/or the parameter values of the questions. The instructor can also limit the view to one question per page for students and prevent them from going back to previous questions. The effectiveness of these options in limiting cheating/ plagiarism— and consequently the reduced need for further proctoring —might have contributed to the high popularity of this assessment method among the faculty. The remaining assessment methods in the decreasing order of their prevalence were project/term paper (50%), semi-online synchronous exam (40%), oral presentation/exam (33%), and semi-online asynchronous exam (28%). Our survey also revealed that 70% of the faculty used the open-book/open-note exam while 33% tried closed-book/closed note exams. The preference of open-book/open-note exam among faculty could be also justified by the decreased need for proctoring tools. In fact, our data (faculty survey question #7) revealed that among those faculty who employed open-book/open-note exam, only 27% used Zoom camera and microphone for proctoring of the exam. 21% used lockdown browsers (e.g. respondus), while 61% did not have any proctoring. However, when the exams were closed-book/closed-note, 56% of the faculty decided to proctor the exam using Zoom camera and microphone, 18% chose to use the lockdown browsers and 35% did not proctor. We also evaluated the association of cheating/plagiarism with various assessment methods by calculating the Pearson correlation of faculty's assessment methods with their trichotomized perception of online cheating (less cheating, the same, more cheating) relative to that of face-to-face (faculty survey question #8). The results revealed no statistically significant correlation between cheating and assessment methods except for the following: Fully online exam (correlation= -0.40, p-value<0.0001), Semi-online asynchronous exam (correlation= 0.26, p-value=0.005), Open-book/Closed-book (correlation= -0.30, p-value=0.002). This data analysis shows that while semi-online asynchronous exams were associated with an increase in the perceived cheating, a fully online or open-book/open-note exams had an association with a decrease in instructor's perception of cheating.

### D. PERCEIVED FACULTY SKILLS THAT NEEDED ENHANCEMENT

Faculty indicated various topics that they were interested to enhance their skills in, as summarized in Table 2.

**TABLE II**
**A list of topics identified by faculty for further skill enhancement. Respondents could choose as many topics as they were interested to learn.**

| Survey Question # | Topics | Percentage of faculty interested |
|---|---|---|
| Question #9 | What are the major requirements of syllabus for an online course? | 38% |
| | Basic BB features: How to create/modify/improve BB for my course | 26% |
| | More advanced BB features: How to create online surveys /discussion groups/quizzes that reduce the potential of cheating, how to automatically export grades to BB gradebook, how to use Master Shell in BB, etc. | 58% |
| | Zoom features (basic and advanced): How to schedule/record a meeting, how to use Zoom's Whiteboard or OneNote, how to do breakout rooms, etc. | 39% |
| | Multimedia skills: How to create interactive multimedia files using Kaltura Capture, Camtasia or Snagit, how to use Alt captions in media you generate (Word, PPT, page in BB) to facilitate accessibility | 39% |
| | Assessment: How to use automatic grading tools (e.g. Gradescope) | 54% |

About 60% of the faculty needed to learn about the advanced features of BB (e.g. how to create online surveys or make quizzes with randomized questions/personalized parameter values). Also, more than half of the faculty were interested in learning about automatic grading tools (e.g. Gradescope).

Close to 40% of the faculty needed to learn how to create a syllabus for an online class or become more competent with using Zoom features. A similar percentage of participants indicated interest in enhancing their multimedia skills (e.g. working with Kaltura Capture, Camtasia or Snagit).

Finally, 26% of the faculty needed more training to become familiar with basic features of BB. In the optional comments (faculty survey questions #10-11), some faculty members expressed their concerns about the delivery of the hands-on components of their courses and requested some general guideline on how to address this issue for an online instruction.

## IV. DISCUSSION

Online learning during COVID-19 pandemic has shown to be challenging especially for vulnerable, disadvantaged, and underrepresented students whose difficulties have been magnified.

As the pandemic continues, a small body of literature on educational impact of COVID-19 is starting to emerge. A group of investigators conducted a U.S. nationwide survey study among faculty and students of STEM fields in June 2020. Their results highlighted the gender disparities in online learning during pandemic: female faculty and students reported more challenges in technological issues and adapting to remote learning compared with their male peers [15]. They also found out that 35.5% of doctoral students, 18.0% of master's students and 7.6% of undergraduate students would have a delayed graduation due to pandemic [14]. Hispanic and Black undergraduates were two times and 1.7 times more likely, respectively, to delay graduation relative to Whites.

Our survey focused on identifying various challenges that engineering faculty and student confronted during the online instruction in Spring 2020. In this section, we will discuss the challenges and propose relevant interventions to improve the online delivery of engineering courses.

### A. STUDENT CHALLENGES

Our results revealed that a quarter of our students did not have access to reliable internet connection, triggering a concern about widening of the digital equity gap among students during COVID-19 pandemic. A successful online education without a reliable and robust technical infrastructure is not possible [17].

With COVID-19 and the abrupt transition to online teaching, access to reliable internet connection and personal computer/tablet have become major factors affecting the learning/teaching outcomes for students and faculty. To address this issue, institution can provide WiFi access on campus's open areas and well-ventilated buildings while monitoring for social distancing and sanitizing the surfaces frequently. For those who require computing devices, a loaner program can be implemented where students can borrow laptops for a certain period of time to access the course materials and complete the course requirements. The institution can also provide a virtual desktop environment for students to access all necessary software. Using free scanning applications on smartphones or tablets can address the lack of access to scanners.

Our survey also indicated that about 30% of engineering students had work-life balance issues, while more than half of students lacked motivation or did not have access to a private space to attend classes. These results are consistent with those reported in a recent study conducted at Biomedical and Chemical Engineering departments of a Hispanic-serving institution [13]. While the percentage of our students who had issues with lack of motivation or private space seemed to be higher, both studies highlight the necessity of providing more socio-emotional support for students during the difficult times of pandemic. Despite some similarities, there are several differences between our study and [13] that need to be pointed out: while their study is based on a qualitative surveying from 170 students within two Engineering departments, our study uses qualitative and quantitative surveying from 627 students and 110 faculty from six Engineering departments. So, our study has a larger cohort size of students from more engineering disciplines. Furthermore, our survey questions were qualitative and quantitative. In addition to students, we also included another important education stakeholder (faculty) in our survey. Finally, while the study of [13] focused on asynchronous online classes, our classes at CSULB were synchronous live.

Students identified various challenges they experienced in online synchronous instruction of courses through Zoom including lack of peer-support/interaction, focus, engagement, and clear guideline from instructors. They also indicated difficulties with time management and Zoom fatigue. Peer-support/interaction has shown to improve the success rate of students especially those from underrepresented groups [18]. Lack of peer-support during the online instruction in the COVID-19 era negatively affects the motivation and learning outcomes of the students. However, the remaining raised issues could be addressed in part by employing appropriate teaching techniques by faculty as follows: breaking down a long lecture into shorter segments with more frequent breaks, encouraging group discussion among students, making themselves available during the exams, providing students with a clear roadmap for the online course, making the recordings of the live lectures available after the lecture is over. The latest would help struggling students to learn at their own pace [13]. To assist with the time management issue, faculty can design practice exams to allow students to familiarize themselves with the questions' setup and adapt with the exam's style before the actual exam.

### B. FACULTY CHALLENGES

Establishment of institutional quality standards related to online education is of paramount importance in online education. Effective communication is the key factor in bridging the divide and reconciling administrator and faculty for an enhanced online education [19]. A considerable number of our faculty reported lack of access to hardware, software and necessary tools for online instruction. Especially, in the absence of traditional in-class whiteboard, many faculty members indicated lacking an online writing tool. This issue can be addressed by institution's budget allocation to acquire necessary hardware and tools (e.g. personal computer/tablet with web camera, active pen for touch screen devices, digital clipboard, document camera).

Development of online learning assessment methods as rigorous as in conventional face-to-face setting to prevent cheating/plagiarism does not seem straightforward [10, 20]. Some of our faculty have used camera/microphone to proctor exams during Spring 2020. However, as our student survey indicated, the use of camera/microphone can raise equity concerns (for those who do not have access to camera and cannot afford it) and privacy concern (for monitoring students' private space). To address these valid concerns, faculty are advised to choose alternative methods for reducing cheating during online exams. Randomizing the exam questions by shuffling both the problem statements and the multiple choices or randomly selecting a subset of questions in a question library with individualized input variables could be practical solutions. Fortunately, most LMS provide these options. However, although 99% of postsecondary US institutions have an LMS in use, only approximately half of faculty at those institutions have been using it on a regular basis [21]. As a result, many faculty members are not familiar with the basic or advances features of their LMS or other tools for effective online instruction. Our survey result confirmed this observation. In fact, our faculty identified a broad range of topics related to BB or other online teaching tools that they felt the need to enhance their skills in. Institutions could address this issue by organizing training workshops, webinars, short-courses, and discussion panels for their faculty to enhance their online teaching skills. At CSULB, stipends were offered in summer 2020 to further incentivize faculty participation in these professional development programs.

Hands-on training is an integral component of engineering education. Following the abrupt conversion of classes to the online format in Spring 2020, many instructors adopted simulations or processing of already acquired data for engineering students to complete their course projects. Our survey indicated the faculty's need to learn about additional effective ways for providing hands-on training/experience. Depending on the content of the course, employment of "home lab kits" and recording of the lab experiments could partially help. However, design, preparation, distribution/collection of the lab kits or recording of the experiments can be extremely time consuming for faculty especially given all the access restrictions to on-campus labs and additional safety precautions imposed by COVID-19 pandemic. Virtual labs might be a more effective solution. Additionally, remotely accessible labs where the experiment setup is on campus and students use tools for remote control and managing of the setup can be employed, whenever possible [13].

*C. SUMMARY OF PROPOSED INTERVENTIONS*

From the analysis of the survey results we propose several intervention strategies that can be employed by stakeholders at different levels to improve the online instruction of engineering courses. The proposed strategies are summarized as follows:

- Strategies for institution/engineering administration
  - Budget allocation to provide basic equipment for the online instruction (e.g., personal computer/tablet preferably with webcam/camera, reliable internet connection) to both faculty and students in need
  - Creating a virtual desktop environment and allowing faculty and students to access necessary software
  - Organizing training workshops for faculty/students to further familiarize with online teaching/learning technology and tools
  - Providing a syllabus template for online courses including all the important information
  - Development and organization of systematic repository of resources pertinent to engineering online instruction
- Strategies for engineering faculty
  - Leveraging on the institution's LMS to manage the course, grades, forum discussions and exams
  - Breaking down a long lecture into shorter segments with more frequent breaks
  - Encouraging group discussion or problem-solving activities among students (e.g. Zoom breakout rooms)
  - Being available during the exams (e.g. on Zoom) to answer students' questions
  - Providing students with a clear roadmap and instruction for the online course.
  - Making the recordings of the live lectures available after the lecture.
  - Administering practice exams for students
  - Using open-book/open-note and synchronous assessment methods (e.g. randomized questions/ restricted time/ question pools on LMS).
  - Avoiding using camera/microphone to proctor exams
  - Employment of "home lab kits", recording of the hands-on experiments and virtual labs to partially address the hands-on training aspect of the course
- Strategies foe engineering Students
  - Using free scanning applications on their smartphones

Most of the proposed solutions were implemented at the CSULB college of Engineering before the beginning of Fall 2020 semester. Our future work will include evaluation of the efficacy of the implemented interventions by conducting a post-intervention survey at the end of Fall 2020 semester.

**V. CONCLUSION**

We conducted an observational study to improve the online instruction of engineering courses during the COVID-19 pandemic by surveying students and faculty at our minority-serving institution. Our surveys identified various logistical, technical, learning/teaching challenges and we proposed strategies to address them. Our future work would focus on evaluating the efficacy of each proposed strategy. We believe that sharing the current results with other engineering educators can aid with further enhancement of online education during the COVID-19 pandemic.

**VI. APPENDIX**

Appendix includes copies of both faculty and student surveys.

*A. FACULTY SURVEY*

1. What is your home department? Select all that apply
   a. BME
   b. CHE
   c. CECEM
   d. CECS
   e. EE
   f. MAE
2. Check all that you had challenges with (e.g. lack of access or difficulty in operations) in transitioning to online instruction in Spring 2020?
   a. Computer and tablet
   b. Mic or headset
   c. Webcam/camera
   d. Scanner
   e. Document camera
   f. Online writing tools (e.g., digital pen)
   g. Printer or cartridge
   h. Access to reliable internet (at least 3 Mbps down, 1- 3 Mbps up)
   i. Software license
   j. VPN and remote access
   k. ATS helpdesk and online support
   l. Others: Please fill in
3. Which of the following did you primarily use to assess your students' learning in Spring 2020? Check all that apply.
   a. Completely online exams (e.g., BeachBoard Quiz)
   b. Asynchronous semi-online exam (download, pen and paper, scan, upload)
   c. Synchronous (live) semi-online exam (download, pen and paper, scan, upload)
   d. Project/term paper

    e. Oral presentation or demonstration
    f. Others: Please fill in

4. What kind of classes will you be teaching in Summer/Fall 2020? Select all that apply.
    a. Lecture only
    b. Lecture with Non-hands-on lab or activity, e.g., simulation, problem solving, etc.
    c. Lecture with hands-on lab or activity
    d. Others (a type of class not listed above, or being the course coordinator): fill in

5. Which of the following will you primarily use to assess your students' learning in Summer/Fall 2020? Check all that apply.
    a. Completely online exams (e.g., BeachBoard Quiz)
    b. Asynchronous semi-online exam (download, pen and paper, scan, upload)
    c. Synchronous (live) semi-online exam (download, pen and paper, scan, upload)
    d. Project/term paper
    e. Oral presentation or demonstration
    f. Others: Please fill in

6. If you had an online exam (semi- or completely), which type was it? Check all that apply.
    a. Open book/notes
    b. Closed book/notes
    c. Exam that requires the use of specific software
    d. Other (fill in)

7. If you had an online exam, how did you proctor it? Check all that apply.
    a. Using Zoom camera and mic on
    b. Using Lockdown browser and Respondus monitor
    c. Used online exams, but did not proctor it
    d. Other (fill in)

8. What is your perception of the extent of cheating/plagiarism in Spring 2020 relative to prior semesters?
    a. Way less
    b. Less
    c. About the same
    d. More
    e. Way more
    f. I do not know/ Did not use online exam
    g. Other (fill in)

9. Indicate all your topics of interest to enhance your skills (by either attending a workshop or watching a webcast).
    a. Online course syllabus: What are the requirements of an online course syllabus
    b. Basic BeachBoard (BB) features: How to create/modify/improve BB for my course
    c. More advanced BB features: How to create online surveys /discussion groups/quizzes that reduces the potential of cheating, how to automatically export grades to BB Grades, how to use Master Shell in BB, etc.
    d. Zoom features (basic and advanced): How to schedule/record a meeting, how to use Zoom's Whiteboard or OneNote, how to do breakout rooms, etc.
    e. Multimedia skills: How to create interactive multimedia files using Kaltura Capture, Camtasia or Snagit, how to use Alt captions in media you generate (Word, PPT, page in BB) to facilitate accessibility
    f. Assessment: How to do automatic grading using software that helps grading (e.g. Gradescope)
    g. Learning Objective (LO): What are student LOs and module LOs? How to align them?
    h. Other: to be filled

10. If you were to be provided with a personal trainer, what are your top two online teaching challenges that you would like the trainer to help you with for teaching your class more effectively in Summer/Fall 2020?

11. Please provide any additional comments here on the challenges you have faced regarding teaching and issues that we might be able to help you resolve in Summer/Fall 2020.

**B. STUDENT SURVEY**

1. What is your major? (select all that apply)
    a. Biomedical Engineering
    b. Chemical Engineering
    c. Civil Engineering
    d. Construction Management Engineering
    e. Computer Engineering
    f. Computer Science
    g. Electrical Engineering
    h. Mechanical Engineering
    i. Aerospace Engineering
    j. Engineering Technology

2. What's your academic level?
    a. Freshman (mainly taking 100-level courses within your department)
    b. Sophomore (mainly taking 200-level courses within your department)
    c. Junior (mainly taking 300-level courses within your department)
    d. Senior (mainly taking 400-level courses within your department)
    e. Graduate level
    f. Other: please fill in

3. Which of the following will be a challenge for you when taking classes in a fully online environment? Check all that apply
    a. No access to a computer
    b. Sharing computer with others
    c. No access to a computer with camera/ webcam
    d. No internet access
    e. No reliable/high speed internet access
    f. No private/quiet space to work in
    g. Working to support myself or my family, therefore not enough time to study

    h. Time management, Lack of motivation
    i. Others: Please fill in
4. Which of the following challenges did you experience with Zoom synchronous (live) classes in Spring or Summer 2020?
    a. Did not have reliable internet connection
    b. Was not able to focus and follow lectures
    c. Felt socially disconnected from peers
    d. Felt Zoom fatigues (overwhelmed with multiple online sessions)
    e. Did not feel engaged
    f. Others: Please fill in
5. Which of the following challenges did you experience with classes taught in online instruction mode in Spring or Summer 2020? Select all that apply
    a. No clear instructions
    b. Lack of communication
    c. Issues with accessing course material
    d. Issues with using technology/software
    e. Navigating BeachBoard to access assignments/exams
    f. Issues with submitting assignments
    g. Others: Please fill in
6. Which of the following challenges did you experience with online exams? Select all that apply
    a. Time management
    b. No access to a quiet space to take the exam
    c. No access to a printer/scanner to print out or submit the exam
    d. No access to a reliable internet
    e. No access to instructors during exams
    f. Exams were more difficult than in-class exams
    g. Issues with methods of proctoring exams, such as: Respondus Lockdown Browser, Zoom, etc
    h. Noticed more cheating among classmates
    i. Others: Please fill in
7. Do you feel comfortable using a cell phone or computer camera and showing your face, only for the purpose of identification during demos, presentations, or exams?
    a. Yes
    b. No
    c. I cannot because I don't have a camera
    d. Other concerns: Please fill in
8. How would you rate your overall experience with taking classes in an online mode of instruction in Spring 2020?
    a. Very satisfied
    b. Satisfied
    c. Neither satisfied nor dissatisfied
    d. Dissatisfied
    e. Very dissatisfied

## VII. ACKNOWLEDGEMENT

This research is partially supported by CSULB Champions program through Coronavirus Aid, Relief, and Economic Security (CARES) Act funding. The authors would like to thank Dr. Daniel Whisler, Dr. Shabnam Sodagari and Ms. Asieh Jalali-Farahani for their help with designing the surveys.

**REFERENCES**

[1] J. Bourne, D. Harris, and F. Mayadas, "Online engineering education: Learning anywhere, anytime," Journal of Engineering Education, vol. 94, pp. 131-146, 2005.

[2] R. M. Lima, P. H. Andersson, and E. Saalman, "Active Learning in Engineering Education: a (re)introduction," European Journal of Engineering Education, vol. 42, pp. 1-4, 2017.

[3] J. L. Bishop and M. A. Verleger, "The flipped classroom: A survey of the research," in ASEE national conference proceedings, Atlanta, GA, 2013, pp. 1-18.

[4] J. E. Mills and D. F. Treagust, "Engineering education—Is problem-based or project-based learning the answer," Australasian journal of engineering education, vol. 3, pp. 2-16, 2003.

[5] S. Asgari, B. Penzenstadler, A. Monge, and D. Richardson, "Computing to Change the World for the Better: A Research-Focused Workshop for Women," in Research on Equity & Sustained Participation in Engineering, Computing, and Technology (RESPECT) Conference, Portland, Oregon, USA, 2020.

[6] S. Asgari and B. Englert, "Teaching Pattern Recognition: A Multidisciplinary Experience," presented at the American Society of Engineering Education (ASEE) Conference- Zone IV, Long Beach, CA, 2014.

[7] P. J. Martínez, F. J. Aguilar, and M. Ortiz, "Transitioning from face-to-face to blended and full online learning engineering master's program," IEEE Transactions on Education, vol. 63, pp. 2-9, 2019.

[8] S. S. Jaggars, "Choosing between online and face-to-face courses: Community college student voices," American Journal of Distance Education, vol. 28, pp. 27-38, 2014.

[9] P. C. Holzweiss, S. A. Joyner, M. B. Fuller, S. Henderson, and R. Young, "Online graduate students' perceptions of best learning experiences," Distance education, vol. 35, pp. 311-323, 2014.

[10] A. Lee-Post and H. Hapke, "Online learning integrity approaches: Current practices and future solutions," Online Learning, vol. 21, pp. 135-145, 2017.

[11] P. Moskal, C. Dziuban, and J. Hartman, "Blended learning: A dangerous idea?, The Internet and Higher Education, vol. 18, pp. 15-23, 2013.

[12] COVID-19 Impact on Education, United Nations Education, Scientific and Cultural Organization (UNESCO), https://en.unesco.org/covid19/educationresponse (Accessed on September 22, 2020).

[13] K. Vielma and E. M. Brey, "Using Evaluative Data to Assess Virtual Learning Experiences for Students

During COVID-19," Biomedical Engineering Education, pp. 1-6, 2020.

[14] G. K. Saw, C. Chang, U. Lomelí, and M. Zhi, "Fall Enrollment and Delayed Graduation Among STEM Students during the COVID-19 Pandemic," Network for Research and Evaluation in Education (NREED) Data brief, https://nreeducation.wordpress.com June 2020.

[15] G. K. Saw, C. Chang, U. Lomelí, and M. Zhi, "Gender Disparities in Remote Learning during the COVID-19 Pandemic: A National Survey of STEM Faculty and Students," Network for Research and Evaluation in Education (NREED) Data brief, https://nreeducation.wordpress.com, August 2020.

[16] 2019 CSULB Institutional Data, California State University, Long Beach, https://www.csulb.edu/institutional-research-analytics (Accessed on September 22, 2020).

[17] S. Gold, "A constructivist approach to online training for online teachers," Journal of Asynchronous Learning Networks, vol. 5, pp. 35-57, 2001.

[18] S. N. Williams, B. K. Thakore, and R. McGee, "Providing social support for underrepresented racial and ethnic minority PhD students in the biomedical sciences: a career coaching model," CBE—Life Sciences Education, vol. 16, p. ar64, 2017.

[19] L. E. Wickersham and J. A. McElhany, "Bridging the divide: Reconciling administrator and faculty concerns regarding online education," Quarterly Review of Distance Education, vol. 11, p. 1, 2010.

[20] J. Baron and S. M. Crooks, "Academic integrity in web-based distance education," TechTrends, vol. 49, pp. 40-45, 2005.

[21] T. Dahlberg, T. Barnes, K. Buch, and A. Rorrer, "The STARS alliance: Viable strategies for broadening participation in computing," ACM Transactions on Computing Education (TOCE), vol. 11, p. 18, 2011.

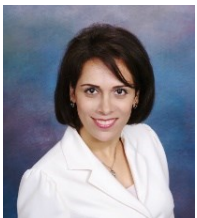

**Shadnaz Asgari** received her Ph.D. in electrical engineering with emphasis on telecommunication and signal processing from University of California (UCLA) in 2008. She was a postdoctoral researcher (2008-2010) and a research faculty (2010-2012) at UCLA Neurosurgery department conducting biomedical signal processing and machine learning for clinical decision support. She joined the faculty of Computer Engineering and Computer Science at California State University, Long Beach (CSULB) in 2012. She is also the Founding faculty and Chair of Biomedical Engineering department at CSULB. Dr. Asgari has been extensively involved with the curriculum development of Biomedical Engineering program at CSULB. Her current research area is computational physiology. Prof. Asgari has more than 80 peer-reviewed research publications. She is also the recipients of several awards including 2010 UCLA Brain Injury Young Investigator Award, 2015 CSULB Early Excellence in Academic Career Award, 2018 Google exploreCSR Award, 2019 Google Faculty in Residence Award and 2020 CSU Faculty Innovation and Leadership Award.

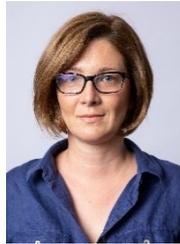

**Jelena Trajkovic** received her PhD (2009) and MS (2003) in information and computer science from the University of California, Irvine, and a Dipl. Ing. degree in electrical engineering from University of Belgrade, Serbia (2000). She was a ReSMiQ postdoctoral scholar at Ecole Polytechnique de Montreal (2010-2012) and an Assistant Professor (2012-2018) and an Affiliate Assistant Professor appointment (2018-2020) in the Electrical and Computer Engineering Department at Concordia University in Montreal. She joined Computer Engineering and Computer Science department at California State University, Long Beach as an Assistant Professor in 2018. Dr. Trajkovic has extensive research experience in the domains of network-on-chip, silicon photonics, multicore systems, parallel applications, and avionics systems. Her research has been recognized by three Best Paper conference Awards, and the Graduate Dean's Dissertation Fellowship at UC Irvine. She is also the recipient of the Teaching Excellence Award at Concordia University (2017).

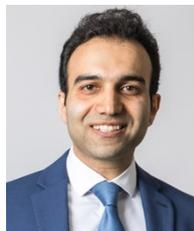

**Mehran Rahmani** received his PhD in structural and earthquake engineering from the University of Southern California (USC) in 2014, MS in electrical engineering from USC in 2013, and MS in structural engineering from Sharif University of Technology (Iran) in 2009. He joined the CSULB Civil Engineering and Construction Management Department as an Assistant Professor in 2017. Prior to joining CSULB, Dr. Rahmani worked as a structural engineer between 2014 and 2017. He is a licensed Professional Engineer (PE) in the state of California. His research focuses on structural system identification, structural health monitoring and earthquake damage detection of buildings using sensory data. His current research is on developing a wave-based methodology for remote post-earthquake damage detection in full-scale buildings and bridges. Dr. Rahmani has authored and co-authored several peer-reviewed technical papers and is an active member of professional societies including ASCE, EERI, and ACI.

**Wenlu Zhang** received her Ph.D. in Computer Science from Old Dominion University in 2016. She was a postdoctoral researcher (2016-2017) in Washington State University. She joined the faculty of Computer Engineering and Computer Science at California State University, Long Beach (CSULB) in 2017. Dr. Zhang's research interests include machine learning, data mining, computational biology and computational neuroscience. Dr. Zhang received an Outstanding Research Assistant Award in Old Dominion University. She and 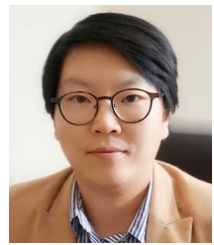 her team participated in the MICCAI Challenge on Circuit Reconstruction from Electron Microscopy Images (CREMI). Her team ranked first on synaptic cleft detection and second on neuron segmentation tasks. She also serves as a program committee for the ACM SIGKDD, ICML, ICLR, MICCAI, CIKM and ICDM.

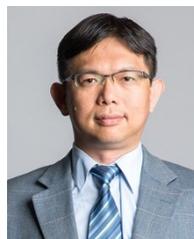

**Roger C. Lo** received his Ph.D. and M.E. from Texas A&M University in 2008 and 2002, respectively, and B.S. from National Chung Hsing University in Taiwan in 1997, all in chemical engineering. He worked as a postdoctoral researcher between 2008 and 2009 in the Department of Chemistry and Biochemistry at California State University, Los Angeles. He joined the faculty of the Chemical Engineering Department at California State University, Long Beach in Fall 2009. Dr. Lo's research interest focuses on microfluidics and its applications at the interface of biology, chemistry, and engineering, such as high-throughput detection and separation of chemical/biological species and modular chemical/biological reactions. His work has been published in various peer-reviewed journals, such as Lab on a Chip, Electrophoresis,

Analytical Chemistry, Soft Matter, and Nature Microsystems & Nanoengineering. Dr. Lo is an active member of AIChE and ASEE.

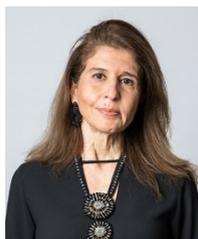

**Antonella Sciortino** received her BS in civil engineering from the Politecnico di Bari, Italy, and her MS and PhD in civil engineering from the University of California, Los Angeles (UCLA). She was a 1996 Fulbright scholar. Dr. Sciortino is a Professor in the Civil Engineering and Construction Engineering Management Department at California State University, Long Beach (CSULB). Currently, she is serving as the Interim Associate Dean for Academic Program for the College of Engineering. Prior to joining CSULB, Dr. Sciortino worked as a visiting researcher at the U.S. Salinity Laboratory in Riverside, California; as a postdoctoral fellow and lecturer at UCLA; and as a visiting researcher and project engineer for Italian government agencies. She holds the Italian Professional Engineering license. Her research focuses on numerical modeling of groundwater and vadose zone flow and contaminant transport processes. Dr. Sciortino is the recipient of numerous recognitions, including the 2006 AESB Faculty of the Year Award, 2009 Northrop Grumman Excellence in Teaching Award, 2010 CSULB Alumni Association Most Inspirational Professor Award, ASCE Orange County Branch 2010 Distinguished Engineering Educator Award, and the 2011 CSULB Distinguished Faculty Teaching Award.